\documentclass{webofc}

\usepackage[varg]{txfonts}   
\usepackage{hyperref}
\usepackage{url}
\hypersetup{colorlinks=true,citecolor=blue,urlcolor=blue,linkcolor=blue}
\usepackage{bm}

\graphicspath{{graphics/}}

\begin{document}
\title{QCD critical point: recent developments}
%
%

\author{\firstname{Mikhail} \lastname{Stephanov}\inst{1,2}\fnsep\thanks{\email{misha@uic.edu}}
}

\institute{Department of Physics, University of Illinois, Chicago, Illinois 60607, USA
\and
       Kadanoff Center for Theoretical Physics, University of Chicago, Chicago, Illinois 60637, USA
          }

\abstract{Recent developments aimed at mapping QCD phase diagram and
  the search for the QCD critical point in heavy-ion collisions are
  briefly reviewed.
}
\maketitle
\section{Introduction}
\label{sec:intro}
\begin{figure}[h]
\centering
\includegraphics[width=10cm,clip]{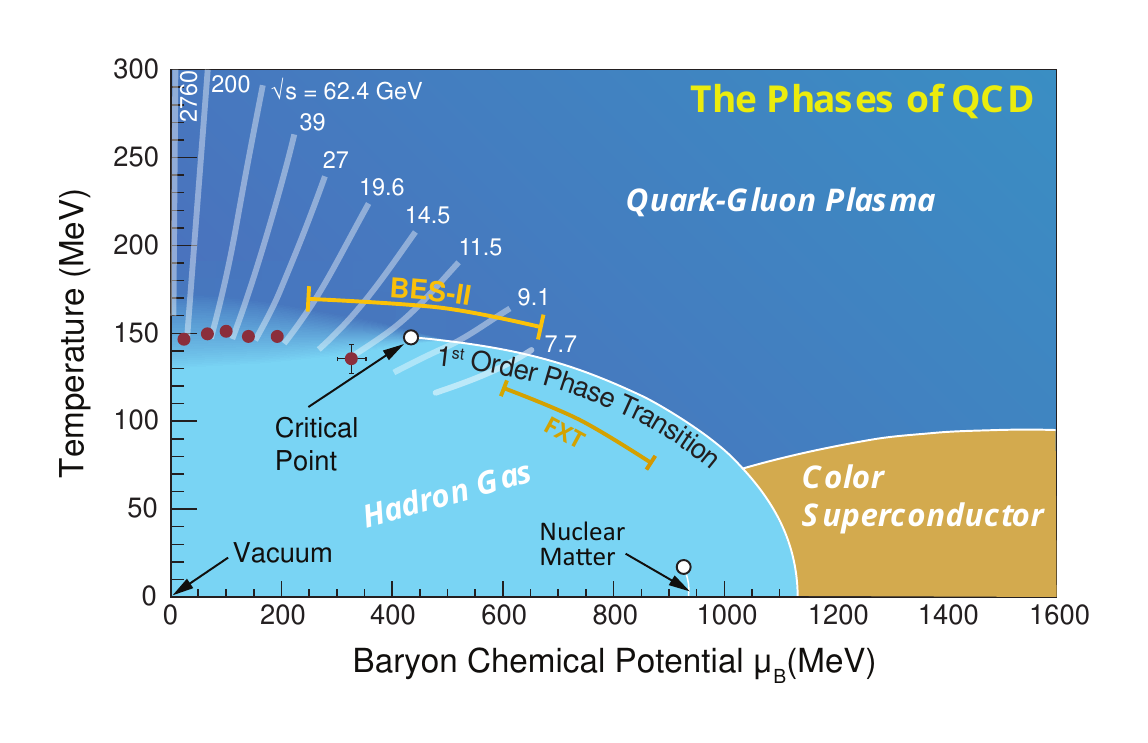}
\caption{QCD phase diagram and beam energy scan experiments,
  Ref.\cite{Du:2024wjm}.}
\label{fig:pd}       
\end{figure}

Mapping the phase diagram of QCD is a fascinating theoretical
challenge with experimental implications ranging from the early
Universe to neutron stars and to heavy-ion collision experiments. This
report focuses on the recent developments in the search for the QCD
critical point -- the point where the crossover at low baryon chemical
potential becomes the first-order phase transition separating the
hadron gas and the quark-gluon plasma phases of QCD. The search for
the critical point is a major component of the heavy-ion collision
experiments at RHIC~\cite{STAR:2010vob,Odyniec:2013BI} (see
Fig.~\ref{fig:pd}). A comprehensive theoretical understanding of
critical point physics and its manifestations in heavy-ion collisions
is crucial for the success of the search. Such a theory effort is
underway, as reviewed, e.g., in \cite{Bzdak:2019pkr,Du:2024wjm}.

Determination of the location of the critical point by a first-principles
theoretical calculation is a longstanding challenge due to the
well-known sign problem in lattice Monte Carlo simulations at {\em finite}
$\mu_B$. The most recent results on this front have been obtained by
using a broad range of theoretical techniques:
Pad\'e resummations of the Taylor expansion in $\mu_B$, whose
coefficients are calculated on the lattice \cite{Clarke:2024ugt}, also
using conformal maps \cite{Basar:2023nkp}, the hybrid lattice plus
gauge-gravity correspondence approach (black-hole engineering)
\cite{Hippert:2023bel} and the functional renormalization group approach
\cite{Lu:2023mkn}.  These are different methods with different
potential sources of systematic errors. It is, therefore, notable, that
 they all point to the existence of the critical point in
the same region of the QCD phase diagram: $T_c\sim 100-110$ MeV and
$\mu_c \sim 420-650$ MeV.

\section{Expected signatures and BES-II results}
\label{sec:expected}

\begin{figure}[h]
\centering
  \begin{minipage}{0.27\textwidth}

\includegraphics[height=.15\textheight]{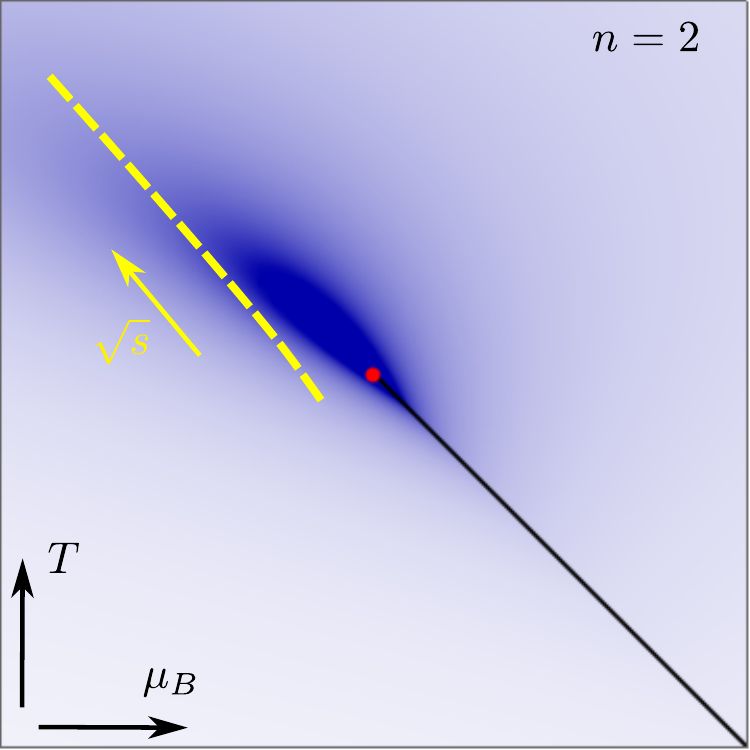}

\includegraphics[height=.15\textheight]{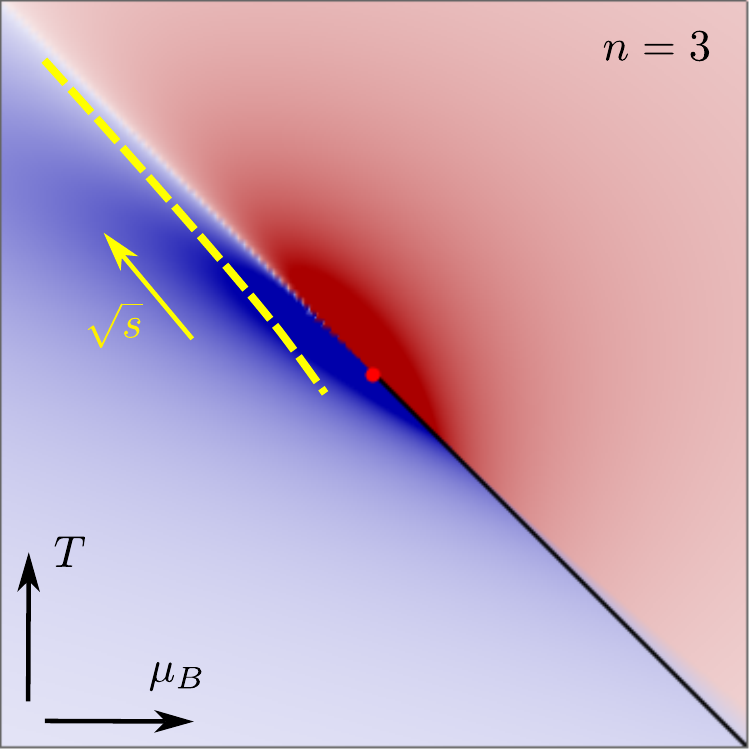}

\includegraphics[height=.15\textheight]{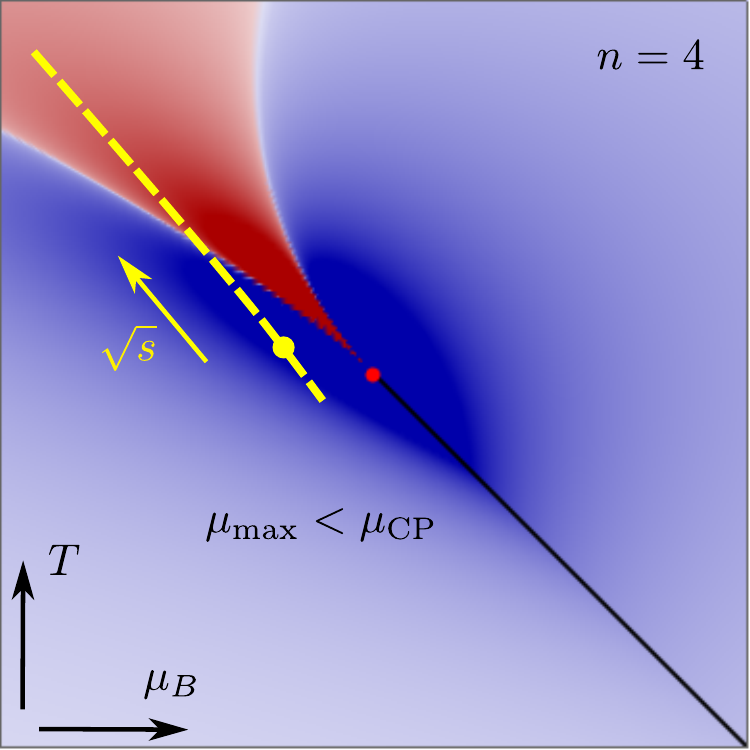}

\end{minipage}
\begin{minipage}{0.25\textwidth}
   
  \includegraphics[height=.14\textheight]{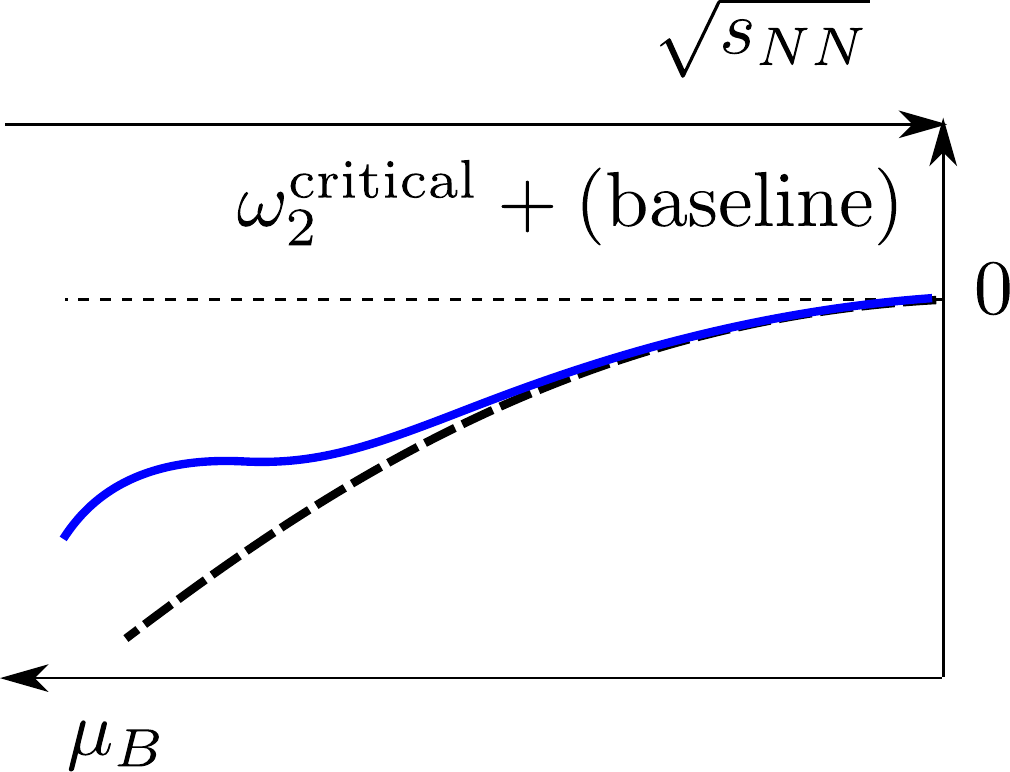}
  
  \includegraphics[height=.14\textheight]{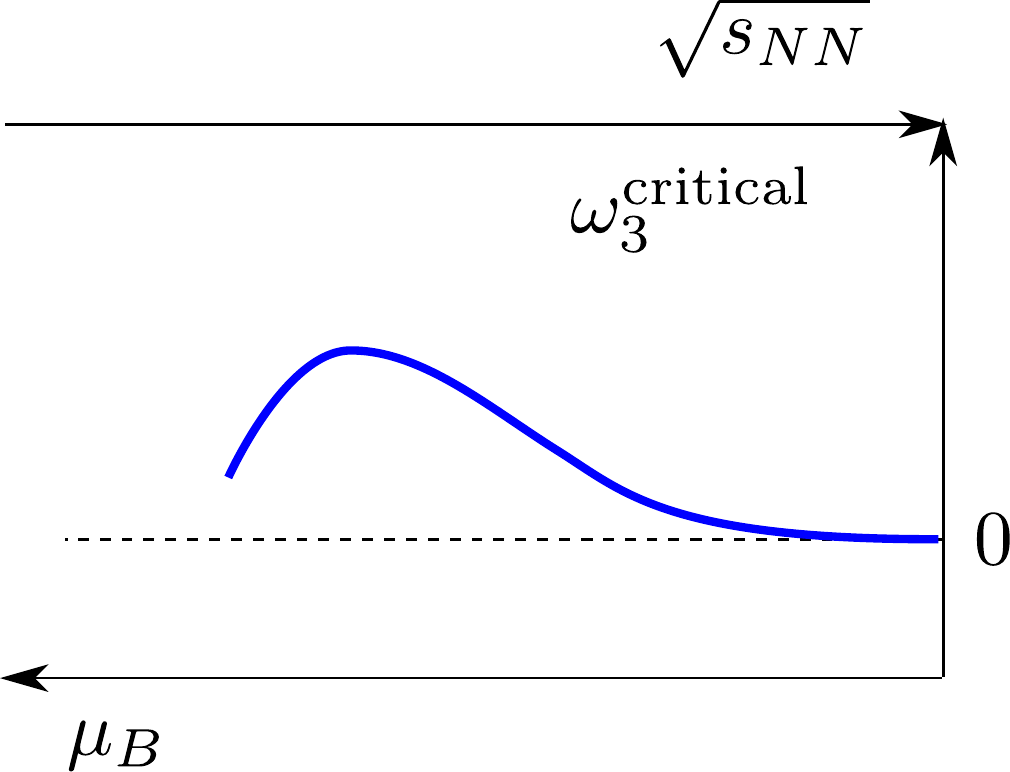}
    
  \includegraphics[height=.14\textheight]{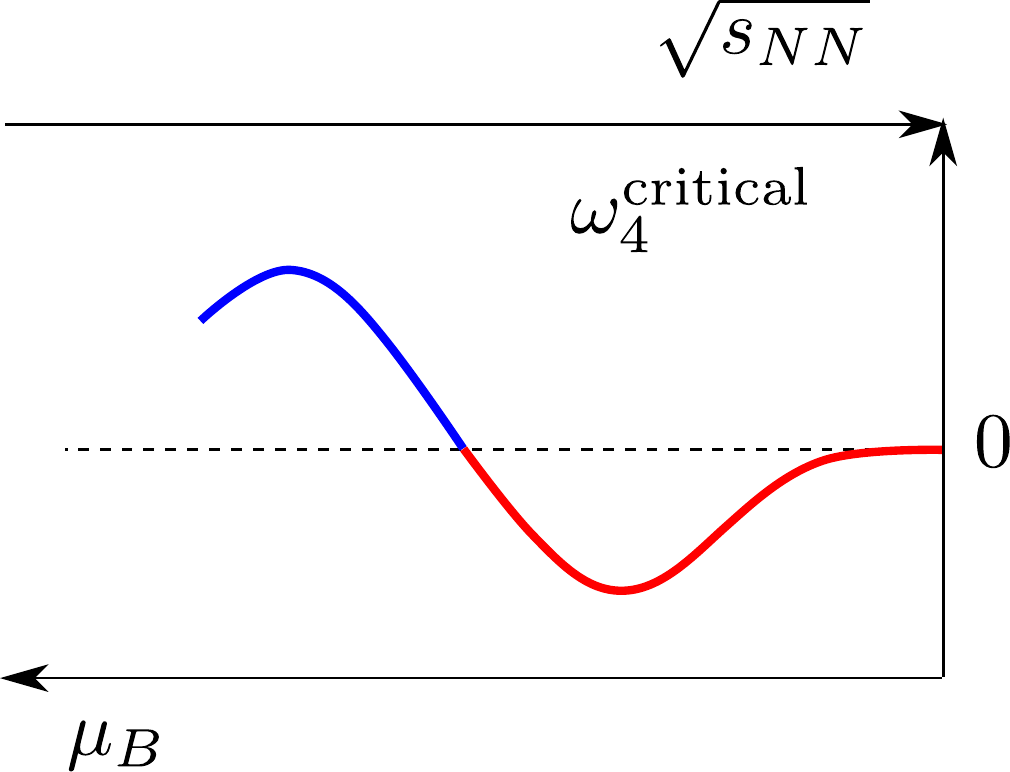}
  
\end{minipage}
\begin{minipage}{0.44\textwidth}
    \vskip 2em
    \hskip 1.5em \includegraphics[height=0.45\textheight]{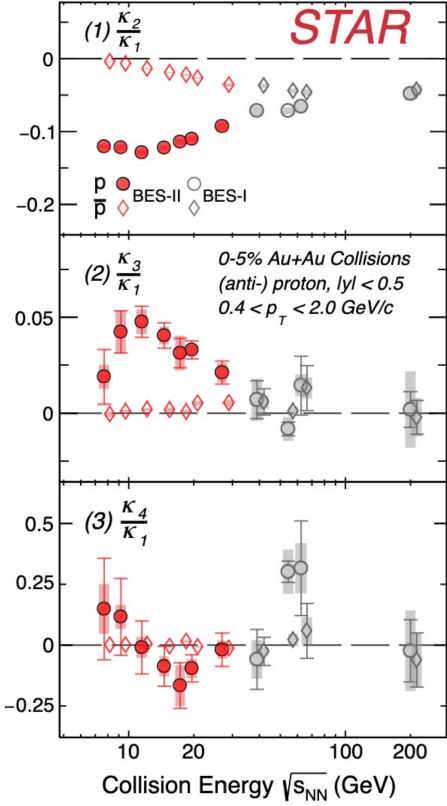}
  \end{minipage}
  \caption{The signatures of the critical point expected in
    experimentally measured factorial cumulants of proton
    multiplicity. Each row in the {\em left column} shows a sketch of
    the QCD phase diagram region near the critical point (red circle) with
    the density plot of the baryon density
    cumulant/susceptibility~$\chi_n$ for $n=2,3,4$ \cite{Bzdak:2019pkr}. The
    cumulants are positive/negative in the blue/red regions. The yellow dashed
    line is the freezeout trajectory -- a set of points where the
    heavy-ion collision fireball freezes out as a function of the
    collision energy $\sqrt s$ (as $\sqrt s$ increases in the direction of
    the arrow). The yellow circle marks the point where the cumulant
    reaches maximum value. The {\em middle column} shows qualitatively the
    corresponding expectation for the collision energy dependence of
    the corresponding (normalized) factorial cumulant of the proton
    multiplicity $\omega_n\equiv\kappa_n/\kappa_1$ due to the critical
    fluctuations. For $n=2$ also the noncritical baseline (evident in the
    experimental data from BES-I and understood as the consequence of
    baryon number conservation \cite{Vovchenko:2021kxx}) is added.  The {\em right column}
    shows experimental results from BES-II at RHIC reported by STAR
    \cite{Pandav:CPOD2024}.  }
\label{fig:signatures}       
\end{figure}

At a recent CPOD 2024 meeting in Berkeley, the STAR collaboration at RHIC
presented the results of the latest beam energy scan (BES-II)
experiment~\cite{Pandav:CPOD2024}. The experiment explored the region
of the phase diagram marked BES-II in Figure~\ref{fig:pd} (where
the collision energies $\sqrt s$ and corresponding fireball expansion
histories are also indicated).  To appreciate the significance of
these results it is helpful to recall the expected signatures of the
critical point, see e.g., review~\cite{Bzdak:2019pkr}.
Figure \ref{fig:signatures} illustrates the expected contribution of
the critical point to the baryon density susceptibility
(left column), reflecting the QCD equation of state, and to the
factorial cumulants of the proton multiplicity fluctuations (middle
column), measured experimentally.  While the approximate
proportionality relationship between the order $n=2$ (Gaussian)
cumulants of these quantities has been known previously, a similar
relationship for higher order cumulants ($n=3$ and 4) has been
established recently via the maximum entropy method in
Ref.\cite{Pradeep:2022eil}, summarized in
Section~\ref{sec:max-entropy}.

The purpose of Figure \ref{fig:signatures} is to highlight the
expected {\em qualitative} features of the {\em non-monotonic}
dependence of these quantities on the collision energy $\sqrt s$. In
summary: one expects a peak (or bump) in the 2nd and 3rd cumulants and
a dip followed by a peak in the 4th cumulant as $\mu_B$ is increased,
i.e., $\sqrt s$ is {\em decreased} \cite{Bzdak:2019pkr}.

The quantitative characteristics of these non-monotonic features --- such as
the position,  height, and width of the peaks --- are sensitively
dependent on the quantitative properties of the QCD equation of
state near the critical point --- such as the location, strength and shape of
the critical point singularity. While qualitative features of this
singularity are universal, the quantitative properties can be
described in terms of the non-universal parameters introduced in
Ref.\cite{Parotto:2018pwx}, see Section \ref{sec:param-eos}.  The
sensitivity to these unknown parameters makes it difficult to predict
the critical point signatures quantitatively~\cite{Karthein:2024ipl}. Conversely, observation
of such signatures would allow us to determine or tightly constrain
these important parameters of the QCD equation of state, which is the
goal of the experiments.

While quantitative comparison to experimental data has to wait for
more quantitative theoretical calculations taking into account not only
the equilibrium expectations, but also important non-equilibrium
effects, it might still be helpful to analyze the qualitative features
of the recent experimental results shown in the {\em right} column of
Figure~\ref{fig:signatures} from the prospective of the equilibrium
theory expectations (the {\em middle} column):

\renewcommand\labelenumi{(\theenumi)}

\begin{enumerate}

\item

  The apparent dominant feature of the measured $n=2$
  cumulant is its monotonic decrease towards lower
  $\sqrt s$ (increasing $\mu_B$). This effect has been already
  observed in BES-I data and attributed to baryon number conservation
  (see, e.g., Ref.\cite{Vovchenko:2021kxx}). What is notable is that
  this monotonic decrease is interrupted at lower collision energies
  $\sqrt s \lesssim 11$ GeV.  Remarkably such an {\em excess} over the
  baseline would be consistent with the presence of the critical point
  as shown in the sketch for $\omega_2$ in the middle column.

\item

  The dominant feature of the measured $n=3$ factorial cumulant is the
  {\em peak} at around $\sqrt s=11$ GeV. This feature is also in
  apparent qualitative
  agreement with the equilibrium expectations for $\omega_3$ shown in
  the middle column.
  
\item

  The dominant feature of the $n=4$ factorial cumulant data is the {\em dip} at
  $\sqrt s \approx 19$ GeV. This feature also appears to be in
  qualitative agreement with the equilibrium critical point
  predictions (middle column) \cite{Stephanov:2011pb,Bzdak:2019pkr,Shuryak:2024yoq}. However, to establish the {\em peak} feature
  expected from the critical point at
  lower $\sqrt s$, the data at $\sqrt s < 7.7$ GeV would be necessary. Such data
  could be provided by the fixed target (FXT) component of BES-II (see
  Fig.~\ref{fig:pd}) and/or
  future dedicated experiments.

\end{enumerate}

One must keep in mind that the {\em non-critical} baseline should be
established in order to reach definitive conclusions about the origin
of the apparent features of the data, such as, e.g., the peak in
$\omega_3$ or the dip in $\omega_4$.%
\footnote{
  The various non-critical baselines presented in Ref.\cite{Pandav:CPOD2024}
  fail to adequately explain the data across all collision
  energies.
  Generally, the non-critical baselines are monotonic, which
  prevents them from fully capturing the non-monotonic features of the
  data.
}
In this regard, the
non-monotonicity is a key property of critical fluctuations,
which distinguishes them from the monotonic baseline contributions.

It is important to highlight that the critical point on the phase
diagram is located at a higher chemical potential $\mu_B$ compared to
the position of the maximum for each of the cumulants:
$\mu_{\rm CP}>\mu_{\rm max}$, as indicated in
Fig.~\ref{fig:signatures}, left column. Therefore, {\em if} the
features of the experimental data discussed above are due to the
critical point, this critical point has to be located at higher
$\mu_B$ than 420 MeV --- the freezeout point at the lower
end of the collision energy range reported in
Ref.\cite{Pandav:CPOD2024}, where the interesting non-monotonic
features of the data discussed above appear.  Notably, such a critical
point location would be also consistent with the recent theoretical
estimates
\cite{Clarke:2024ugt,Basar:2023nkp,Hippert:2023bel,Lu:2023mkn}
briefly discussed in Section~\ref{sec:intro}.

It is important to remember that the expectations discussed
above are based on the assumption of (local) thermal equilibrium of
critical fluctuations. Non-equilibrium effects are important in
heavy-ion collisions, of course, and particularly so for determining
the critical fluctuation signatures, see e.g.,
Ref.\cite{Mukherjee:2015swa,Pradeep:2022mkf}. The
recent developments in describing the non-equilibrium evolution of
fluctuations are reviewed in Ref.\cite{Du:2024wjm,Stephanov:2024mdj}
and, briefly, in Section~\ref{sec:hydro-fluctuations}.

The connection
between the QCD phase diagram and experimental observables runs
through hydrodynamic evolution with a given 
equation of state, which includes also the evolution of fluctuations,
followed by freezeout to convert these hydrodynamic fluctuations into
the observable fluctuations of particle multiplicities. The following
sections highlight recent advances along this outline.

\section{Equation(s) of state of QCD for hydrodynamic calculations}
\label{sec:param-eos}

Since the equation of state of QCD in the finite $\mu_B$
region where one searches for the critical point  is not known, it is
necessary to have an efficient parametrization of possible such
equations of state, which could then be discriminated by experimental
data. Some information about the equation of state is available from
lattice calculations, but this information is limited to the region of
small $\mu_B$, typically $\mu_B\lesssim 2-3T$ by the estimated convergence
radius of Taylor expansion. On the other hand, universality of
critical singularity constrains the equation of state near the
critical point itself up to a few unknown non-universal parameters. Such
parameters were introduced and standardized in
Ref.\cite{Parotto:2018pwx}, where a parametric
family of equations of state, which
match the lattice data at $\mu_B=0$ and include a universal
critical point singularity, was constructed.

In a more recent work, Ref.\cite{Kahangirwe:2024cny}, a significantly
improved parametrization of the QCD equation of state was
developed. Its main additional ingredients include the use of the $T'$
expansion scheme, introduced in Ref.\cite{Borsanyi:2021sxv}.
This scheme amounts, essentially, to
reorganization of the Taylor expansion in powers of $\mu_B$ from being
an expansion at fixed $T$ into an expansion at fixed $T'(\mu_B,T)$, where
$T'(\mu_B,T)=T_c$ is the equation for the crossover line. In addition,
the mapping of $T$ and $\mu_B$ on the QCD phase diagram to $r$ and $h$
parameters on the Ising model phase diagram takes into account the
$C$-symmetry $\mu_B\to-\mu_B$:
\begin{equation}
\label{eq:quadmap}
    \frac{T'(\mu_B,T)-T_c}{T^{'}_{T}T_c}
    = - h w \frac{\sin(\alpha_{1}-\alpha_2)}{\cos\alpha_1}\,;\quad
    \frac{\mu_B^2-\mu^2_{B,c}}{2\mu_{B,c} T_c}
    =-w(r\rho\cos\alpha_1
    +h\cos\alpha_2)\,.
\end{equation}
where the parameters, introduced in
Ref.\cite{Parotto:2018pwx}, control the location ($T_c$, $\mu_{B,c}$),
the strength ($w$) and the shape ($\rho$, $\alpha_{1,2}$) of the
otherwise universal critical point singularity.

As a result, the parametric equation of state proposed in
Ref.\cite{Kahangirwe:2024cny} can be used for hydrodynamic
calculations in a wider region of the QCD phase diagram, importantly,
including $\mu_B$ up to 700 MeV, and for a wider range of the critical
point parameters.

\section{Hydrodynamic trajectories near QCD critical point}

Hydrodynamics is a very universal theory which describes a wide range
of physical systems from astrophysical, such as neutron stars, to
subatomic, such as heavy-ion collisions. Given the equation of state
and initial conditions it governs the evolution of hydrodynamic
variables -- conserved densities characterizing the local (near)
equilibrium state of the fluid. Since mapping the QCD phase diagram
and determining the equation of state is the goal of heavy-ion
collisions, hydrodynamics is the primary tool used to
connect the experimental measurements to the underlying QCD equation
of state -- see Ref.\cite{Du:2024wjm} for a recent review.

In order to predict the signatures of the critical point taking into
account non-equilibrium effects it is
crucial to understand the trajectories on the phase diagram traced by
expanding and cooling fireball during its evolution towards
freezeout. Under often used, and reasonable, zeroth order approximation of
ideal hydrodynamics these trajectories follow the lines of constant
specific entropy $\hat s\equiv s/n$, where $s$ is the entropy density
and $n$ is the baryon density. The singularity of the equation of
state near the critical point is responsible for an intricate behavior
of these trajectories, which has been understood and classified
recently in Ref.\cite{Pradeep:2024cca}.

\begin{figure}[h]
  \centering
  \includegraphics[width=12em]{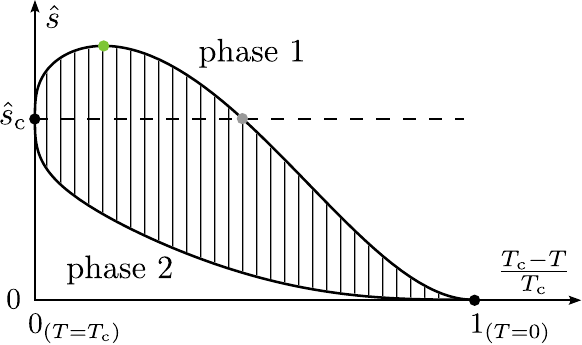}
  \hskip 3em
  \includegraphics[width=17em]{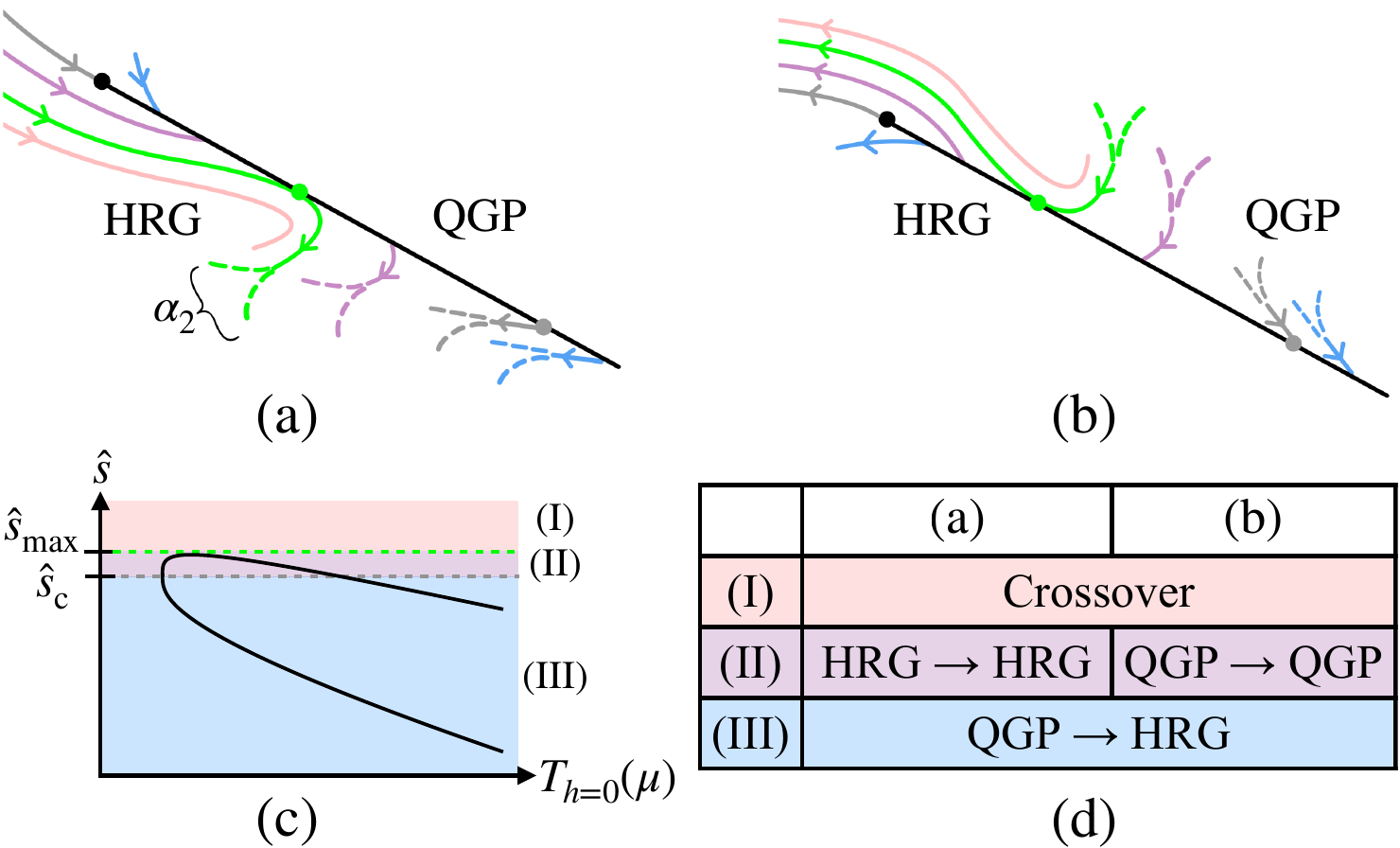}

  \caption{{\em Left:} A sketch of the specific entropy as a function of
    temperature along the phase coexistence
    line. {\em Right:} The classification of isentropic
    trajectories according to the way they behave in the vicintiy of
    the critical point. See Ref.\cite{Pradeep:2024cca}.}
  \label{fig:drop}
\end{figure}

The key to this behavior is the non-monotonicity of $\hat s$ on one of
the branches of the coexistence line (first-order phase
transition). The non-monotonicity is a robust consequence of the
universality of the singularity at the critical point, which requires
$\hat s$ to increase away from the critical point, combined with the
third law of thermodynamic, which requires $\hat s$ to vanish at zero
temperature, as illustrated in Fig.\ref{fig:drop} (left). This gives
rise to an interesting pattern of the trajectories near the QCD
critical point shown in Fig.~\ref{fig:drop} (right).
The unusual (class~II) trajectories enter and reemerge on the {\em same}
side of the coexistence line
    \cite{Pradeep:2024cca}. Possible experimental
    consequences of such a behavior are under investigation.

\section{Hydrodynamic evolution of fluctuations}
\label{sec:hydro-fluctuations}

{\em In equilibrium}, fluctuations of (local) thermodynamic variables are
completely determined by equation of state. In particular, the
universal singular behavior of fluctuations near a critical point
allows us to search for the QCD critical point in
experiments. However, hydrodynamic evolution is characterized by
a sustained deviation of the expanding fireball from equilibrium. These
deviations are manifested in the departure of the fluctuation measures
from their equilibrium values. The evolution of these measures has
been a subject of recent research whose progress is
reviewed, e.g., in Refs.\cite{Du:2024wjm,Stephanov:2024mdj}.

A major recent development is the derivation of the
deterministic equations for the evolution of the {\em non-gaussian}
fluctuation correlators, which are crucial for the QCD critical point
search~\cite{An:2020vri,An:2022jgc}.

Fluctuations are characterized locally by the $n$-point correlators
$H_n\equiv\langle\delta\psi(x_1)\dots\delta\psi(x_n)\rangle$ of
hydrodynamic variables, $\psi$. The evolution equations for these
correlators can be derived from stochastic hydrodynamics. The
equations can be expanded in and organized by the powers of the small
parameter of hydrodynamics -- the ratio of the microscopic and
hydrodynamic scales -- which controls the smallness of the
fluctuations.  In the diagrammatic representation introduced in
Ref.\cite{An:2020vri} this is a loop expansion. The lowest order, tree
level equations were derived in Ref.\cite{An:2020vri}. Loop
corrections lead to the renormalization of hydrodynamic parameters and
the so-called ``long-time tails''.


The quantities particularly suited for writing and solving
fluctuation equations are the Wigner transforms of the
correlators $H_n$, which require a generalization of the well-known
two-point Wigner transform to $n>2$-point correlators. Such a
generalization, $W_n(x,\bm q_1\ldots\bm q_n)$, was introduced in
Ref.\cite{An:2020vri}.  Physically, $W_n$ describes the magnitude
of the correlations of fluctuations with wave-vectors
$\bm q_1,\ldots\bm q_n$ (in the local rest frame of the fluid at point
$x$).

\section{Freezeout and maximum entropy}
\label{sec:max-entropy}

The correlators of hydrodynamic fluctuations
$H_n=\langle\delta\psi(x_1)\dots\delta\psi(x_n)\rangle $ discussed in the
previous section cannot be measured directly. Instead, experiments
measure the correlators of particle multiplicity fluctuations
characterized by the event averages
$G_n\equiv\langle\delta f(x_1,\bm p_1)\dots\delta f(x_n,\bm
p_n)\rangle$ (with
space-time variables $x$ integrated over the freezeout surface), where
$f(x,\bm p)$ are the  
particle phase space distribution functions.

The measured particle correlators $G_n$ are related to the
hydrodynamic correlators $H_n$. In particular, certain integrals of
the particle correlators over momentum space variables
$\bm p_1\ldots\bm p_n$ are constrained by local conservation laws (of
energy, baryon number, etc.) to be equal to the corresponding
hydrodynamic correlators.  However, of course, there is not enough
information in the conservation law constraints alone to determine the
particle correlators $G_n$ from given hydrodynamic correlators~$H_n$.

A solution to this problem has been found recently in
Ref.\cite{Pradeep:2022eil} by applying the maximum entropy method
(MEM). By requiring the entropy of the resonance gas with given
correlators $G_n$ to be maximized under the conservation law
constraints one obtains a unique solution.

Such a solution passes several nontrivial checks. In particular, applied
to the case $n=1$ it reproduces the well-known and time-tested
Cooper-Frye method of freezeout. For critical fluctuations, the MEM
matches the most singular contribution to $G_n$, induced by fluctuations of the
effective critical field $\sigma$ in an earlier approach
\cite{Stephanov:2008qz,Stephanov:2011pb,Bzdak:2019pkr,Pradeep:2022mkf}.
MEM goes further, however, and can describe also subleading and
even non-critical contributions to multiplicity fluctuations within the same
formalism.

In a nutshell, the key observation from the MEM is that the deviations
of the hydrodynamic fluctuations from an uncorrelated ideal resonance
gas correspond directly to the {\em factorial} cumulants of the particle
multiplicities.%
\footnote{More precisely \cite{Pradeep:2022eil},
  the MEM matches irreducible relative
  correlators (IRCs) for particle multiplicities, $\hat\Delta G_n$,
  where lower order correlations are subtracted, to similarly
  constructed hydrodynamic IRCs, $\hat\Delta H_n$. The factorial
  cumulants, $\kappa_n$,  up to
  usually negligible quantum statistics effects, are equal to
  the phase-space integrals of $\hat\Delta
  G_n$.}  This
observation highlights the importance of {\em factorial} cumulants as
experimental fluctuation measures since these cumulants are more
directly related to the QCD equation of state than the normal
cumulants.%
\footnote{The importance of factorial cumulants (FC), as compared to
  normal cumulants (NC), has been understood before
  \cite{Ling:2015yau,Bzdak:2016sxg}. In particular, FCs measure
  deviations of the fluctuations from a Poisson distribution -- the
  natural baseline for {\em discrete} variables such as
  multiplicities.  These deviations are due to ``genuine''
  (irreducible) correlations between the particles. Finally, the
  acceptance dependence of FCs (power scaling with $\Delta y$) is
  much simpler than that of NCs. }

\section{Conclusions and outlook}

The release of the BES-II data by STAR represents a major step towards
uncovering the structure of the QCD phase diagram. It is remarkable
that the non-monotonic features of the data are in qualitative
agreement with the expectations from equilibrium thermodynamics near
the QCD critical point, if one assumes such a point is located at
$\mu_B\gtrsim 420$ MeV. Such a location of the critical point would be
consistent with recent estimates from various theoretical approaches.
A comprehensive quantitative comparison of theory with experimental
data necessitates dynamic simulations of heavy-ion collisions,
incorporating critical fluctuations, which are currently under
development.

This work is supported by the U.S. Department of Energy,
Office of Science, Office of Nuclear Physics Award
No. DE-FG0201ER41195.

%
\bibliography{refs,fluctuations}

\end{document}